\documentclass[11pt]{article}

\usepackage[utf8]{inputenc}
\usepackage[T1]{fontenc}
\usepackage[english]{babel}
\usepackage[margin=1in]{geometry}
\usepackage{graphicx}
\usepackage{tabularx}
\usepackage{booktabs}
\usepackage{array}
\usepackage{ragged2e}
\usepackage{enumitem}
\usepackage{amsmath}
\usepackage[hidelinks]{hyperref}
\newcolumntype{L}{>{\RaggedRight\arraybackslash}X}

\graphicspath{{figures/}}

\begin{document}

\begin{center}
{\Large\bfseries Hybrid Algorithmic Governance in U.S. Welfare Administration:\\
State- and County-Level AI as a Case of Support--Control Convergence\par}

\vspace{1.2em}

{\large Maxim Dedyaev}\\[0.4em]
National Research University Higher School of Economics, Moscow\\[0.2em]
\url{https://orcid.org/0009-0002-2675-5033}\\[0.2em]
E-mail: \href{mailto:maxim.ai.policy@gmail.com}{maxim.ai.policy@gmail.com}
\end{center}

\vspace{1em}

\begin{center}
{\bfseries Abstract}
\end{center}

\begin{quote}
\noindent
This article examines the institutional conditions under which artificial
intelligence systems in U.S. welfare administration come to operate as
instruments of support or as instruments of control. Rather than asking what
welfare algorithms ``really'' are (tools of proactive assistance or
infrastructures of surveillance) the article starts from the premise that
support and control are co-present within the same system, while their
relative balance shifts over time. This movement is conceptualized through
the notion of support--control convergence and the model of an institutional
ratchet. Routine budgetary and political pressures make control-oriented
effects easily measurable and politically capitalizable, whereas a return
toward support requires external intervention of disproportionate force, such
as judicial compulsion, legislative prohibition, or public scandal.
Empirically, the article draws on process tracing of six state- and
county-level cases: NYSDOL fraud detection, Michigan MiDAS, Illinois Medicaid
managed care, LA County homelessness prevention, the Allegheny Family
Screening Tool, and Washington Foster Care. The findings show that the
system's orientation is shaped by institutional design, with the decisive
parameter being the side on which the costs of algorithmic error are placed.
Drift toward control is routine, while reversal is exceptional and costly. In
the MiDAS case, activation required a single administrative decision, whereas
reversal took nine years and a \$20 million settlement; even then, the system
did not return to a support-oriented configuration.
\end{quote}

\vspace{0.5em}

\noindent\textbf{Keywords:} algorithmic governance; welfare administration;
artificial intelligence; support--control convergence; public administration;
process tracing; false-positive costs; administrative discretion;
U.S. federalism

\vspace{1.5em}

\section{Introduction}

The use of artificial intelligence in public administration is no longer
merely a matter of accelerating discrete services. It now concerns the
architecture of state action itself. Within the logic of the ``third wave''
of digital-era governance, algorithmic systems have become a central
component of public administration, redistributing discretion,
responsibility, and power among public officials, models, and private
technology vendors (Dunleavy \& Margetts, 2025; Wirtz, Weyerer \& Geyer,
2019). This shift bears directly on decisions about who gains access to state
resources and who remains outside them. The effects of such systems therefore
depend on how institutions reorganize decision-making procedures from within
(Zuiderwijk, Chen \& Salem, 2021).

This is most visible in welfare administration, where two almost
non-overlapping traditions have emerged for interpreting the same
technologies. The critical tradition shows that predictive systems reproduce
and intensify inequality by operating as an apparatus of discipline and
surveillance (Eubanks, 2018; Citron, 2008; Gilman, 2020; Alston, 2019). The
managerial tradition, rooted in the idea of the proactive state, sees the
same technologies as tools for early risk detection, cost reduction, and
expanded access to services (Margetts \& Dorobantu, 2019; Scholta et al.,
2019; Madan \& Ashok, 2023).

In this literature, such polarization is often framed as a dispute over what
welfare algorithms are: instruments of care or instruments of control. This
article starts from a different premise. What diverges is not merely how
these systems are interpreted, but how the systems themselves develop over
time. The same class of predictive model, operating in the same sector and at
the same level of government, can produce opposite outcomes. The homelessness
prevention system in Los Angeles County, subjected by its operators to
independent experimental evaluation and shown in preliminary assessments to
have substantially reduced homelessness, deliberately places the costs of
error on the state. Michigan's MiDAS system, designed to detect fraud in
unemployment benefits, falsely accused tens of thousands of recipients and
was partially dismantled only after years of litigation.

If the same class of model, in the same sector and at the same level of
government, can generate opposite outcomes, then the question of what a
welfare algorithm ``is'' has been framed incorrectly. Support and control are
co-present within the same system, while their relative balance shifts over
time under institutional conditions.

This article builds on a comparative analysis that mapped thirty cases of AI
deployment in the United States across three levels of public authority and
identified a stable pattern of functional differentiation: control at the
federal level, support at the municipal level, and a ``hybrid'' state level
in between (Dedyaev, 2026). Yet the hybrid category remained the least
developed. It functioned as a residual category and treated the dual
character of the state level as static. Two questions were left unexplained:
the mechanism through which support is converted into control, and the
possible asymmetry of that transition.

It makes a distinct contribution by theorizing how algorithmic support turns
into control. In doing so, it converts a static typology into a dynamic
mechanism.

The fusion of care and control, however, predates algorithms. Historically,
the welfare state has combined assistance with the regulation of poverty
(Piven \& Cloward, 1971). Algorithms do not create this feature; they
automate it, stripping it of its earlier flexibility.

This leads to the main research question: under what institutional conditions
do algorithmic systems in U.S. state and county welfare administration
acquire a dominant orientation toward support, and under what conditions do
they acquire a dominant orientation toward control?

A derivative question follows: is the transition between these orientations
symmetrical, or is there an asymmetry whereby systems drift from support
toward control more easily than in the opposite direction?

Theoretically, the article aims to develop the concept of
\textbf{support--control convergence} as a mode of algorithmic governance
characteristic of a federal system, and to explain the institutional
mechanism of directional drift through the model of an \textbf{institutional
ratchet}. Empirically, it compares six state- and county-level welfare cases
through process tracing -- NYSDOL fraud detection, Michigan MiDAS, Illinois
Medicaid managed care, LA County homelessness prevention, the Allegheny
Family Screening Tool, and Washington Foster Care, in order to trace which
institutional parameters shape the system's functional orientation and the
direction of its movement. The model is grounded in the specific features of
American federalism: decentralization, institutional pluralism, and unequal
administrative capacity across states. The task is to reconstruct a recurring
causal mechanism, not to provide a representative map of the field.

\section{Literature Review}

The literature on AI in welfare administration is organized around a set of
debates in which the central concepts of this article are defined. There are
four such debates: the nature of algorithmic welfare, understood either as
discipline or as efficiency; the form in which it should be described, as
classification or as process; the locus of the explanatory mechanism, in
technology or in institutions; and the directionality of change, as symmetry
or as a ratchet. Each debate ends with an analytical decision that is carried
forward into the empirical part of the study.

\subsection{The Algorithmic State as the Point of Departure for Welfare AI}

The starting point is the broad consensus in the literature on the digital
transformation of governance. Algorithmic systems have moved from the
periphery of automation into the core of administrative decision-making,
redistributing discretion between human actors and infrastructures (Dunleavy
\& Margetts, 2025). This literature also identifies a characteristic
imbalance: the diffusion of AI in the public sector has been studied far more
extensively than the dependence of its effects on institutional context.
Research agendas explicitly identify this gap as a priority (Zuiderwijk, Chen
\& Salem, 2021; Madan \& Ashok, 2023).

This premise leads to a key assumption: the function of an algorithmic system
is an institutional outcome, not a technical property. The question of what a
welfare algorithm ``is'' should therefore be reformulated as a question of
what makes it become what it becomes.

\subsection{Discipline or Efficiency}

The critical tradition is organized around the claim that predictive systems
in the social domain inherit and intensify the power asymmetry between the
state and the poor. Eubanks (2018) describes the ``digital poorhouse'' as a
regime in which automated applicant screening, risk scoring, and interagency
data sharing form an infrastructure of surveillance. Gilliom (2001)
identifies the same logic in the administrative recordkeeping of welfare
recipients, which has always operated as a form of observation and has always
generated quiet practices of resistance. The legal strand of this tradition
documents how automation erodes procedural safeguards (Citron, 2008; Gilman,
2020). The report of the UN Special Rapporteur warns of the dangers of the
``digital welfare state,'' in which social protection becomes an appendage of
control (Alston, 2019). European work in political economy further notes that
the datafication of social protection is accompanied by its ``quiet
marketization'': welfare procedures are increasingly co-produced, while the
quality of the procedure itself becomes part of the welfare good at stake
(Allhutter et al., 2024).

The managerial tradition, by contrast, reads the same technologies in the
opposite register. The call to ``rethink government with AI'' (Margetts \&
Dorobantu, 2019) represents the most explicit pole of this view. Its
theoretical limit, however, is the model of a proactive state that delivers
services before citizens have to request them (Scholta et al., 2019), that
is, a pure configuration of support.

A systematic review of AI adoption in public administration shows that the
field mainly treats the technology as a resource for efficiency,
personalization, and service quality (Madan \& Ashok, 2023). Empirical
studies point to a similar bias. When implemented systems are compared with
public management paradigms, AI aligns most clearly with the values of New
Public Management: economy, performance, and measurability. It aligns much
less often with participation and transparency (Murko, Bab\v{s}ek \&
Aristovnik, 2024). In practice, this view often takes the form of a playbook.
It assumes that competent implementation management can keep the system
within its original purpose (Hysen, 2026).

The two traditions begin to meet when the focus shifts from technology to
organization. Kuziemski and Misuraca (2020) identify a double task for public
administration in the age of AI. The state governs through algorithms, but it
must also govern the algorithms themselves. A study of the Dutch social
insurance system shows the same tension. The same organizations are expected
to serve citizens and pursue fraud at the same time. The balance between
these duties is not stable. It shifts with the political context (Segeren,
van der Voort \& Dobbe, 2025).

Historical sociology of welfare points in the same direction. The duality of
assistance and discipline is constitutive of the welfare state. Welfare
benefits have historically regulated poverty: they expanded in periods of
instability and tightened in periods of order (Piven \& Cloward, 1971). In
the neoliberal era, this dual regulation was redistributed between the social
and penal arms of the state (Wacquant, 2009). The implication is
straightforward. If both traditions describe empirically real configurations,
and if welfare has always contained both assistance and discipline, then the
unit of analysis should be the algorithm as a carrier of both functions. What
is needed is an observable indicator of their current balance. In this
article, that indicator is the distribution of false-positive costs. In a
support-oriented configuration, the state bears those costs. In a
control-oriented configuration, they are shifted onto the recipient.

\subsection{Classification or Process}

The dominant way of describing this duality is through classification. In the
typology developed in the previous study, the state level was assigned to a
``hybrid'' category (Dedyaev, 2026). This solution fit the empirical evidence
but remained analytically incomplete. The hybrid category was defined by what
it was not. The same logic underlies models of layered governance paradigms,
where Weberian, managerial, and participatory elements coexist like
geological strata (Olsen, 2010; Murko et al., 2024). What these approaches
share is a static view of the phenomenon.

The only well-established dynamic alternative is the concept of function
creep. It refers to the gradual expansion of a system beyond its original
mandate without an explicit decision, making the process difficult to detect
or control (Koops, 2021). This concept captures an important feature of
algorithmic governance. Function creep is the outward expansion of a single
function: the system begins to do more than it was originally designed to do.
Convergence, by contrast, refers to the coexistence of two functions whose
relative balance changes over time. The system continues to perform the same
overall role, but it does so differently and with a clear shift toward one
functional orientation. The concept proposed here addresses what the
framework of function creep leaves unexplained: directional change within an
otherwise unchanged institutional purpose.

The analytical conclusion follows directly from this distinction. The unit of
analysis is not the system itself, but its state at a particular point in
time. Drift is therefore understood as a change in the system's profile
across successive states.

\subsection{Technology or Institutions}

As argued above, what a system \textit{is} should be understood as its
institutional outcome. Yet the question of which institutions produce that
outcome remains open. The literature points to three possibilities.

The first is the locus of discretion. Lipsky (1980) located actual social
policy at the ``street level.'' Frontline officials determined what welfare
became for a particular person by allocating attention, leniency, and
judgment. Bovens and Zouridis (2002) showed that informatization shifts this
discretion toward system architects: once a rule is encoded, it no longer
remains open to situational judgment. The literature on digital discretion
points in the same direction (Busch \& Henriksen, 2018).

The implication is important: encoded discretion is asymmetric. A frontline
official could soften a rule in an individual case. A system-level rule, by
contrast, can be softened only by redesigning the system itself. The human
oversight requirements on which much practice-oriented literature relies
(Hysen, 2026) do not remove this asymmetry. Green's (2022) survey of
forty-one policies prescribing human oversight of government algorithms shows
that such oversight often fails to provide meaningful control. Rather than
correcting the system's output, the human reviewer may end up legitimizing
it.

The second possibility is the measurement regime. The sociology of
quantification has shown that public measures do not simply reflect the
worlds they measure. They also reshape them. Measurement is reactive: it
changes the behavior of those being measured (Espeland \& Sauder, 2007).
Target-based governance produces a similar reversal. Instead of measuring
what matters, organizations begin to treat as important what can be measured
(Bevan \& Hood, 2006). This creates an asymmetry between support and control.
The effects of control (money saved, violations detected) are countable,
attributable, and politically usable within a budget cycle. The effects of
support (harm prevented, crises that did not occur) are much harder to
present. The asymmetry of measurability is therefore not only a technical
problem. It is a property of the relationship between metrics and the
political cycle.

The third possibility is the institution of contestation. The framework of
technological due process has shown that automation tends to erode
constitutional safeguards within administrative procedure by default, and
that restoring them requires intentional design (Citron, 2008). The practice
of legal assistance for welfare recipients confirms this point. In automated
welfare, a person's fate often depends on the quality of the channels through
which decisions can be contested (Gilman, 2020). This creates a double
institution. A system with an effective appeal procedure is forced to
internalize part of the costs of its own errors. It is also the only built-in
channel of reversal. All other routes back toward support are external.

A related and equally important line of work comes from system safety. It
treats algorithmic harm as an emergent property of a multi-actor
sociotechnical system (Segeren et al., 2025). This literature addresses a
synchronic task: it maps the risks of the current configuration. The task of
this article is different. It reconstructs how configurations shift over
time.

\subsection{Path Dependence and the Institutional Ratchet}

At this point, it is important to consolidate the theoretical ground
established in the preceding sections:

\begin{enumerate}[itemsep=2pt]
\item the functions of support and control are co-present (Section~2.2);
\item their relative balance is unstable and subject to change
      (Section~2.3);
\item that balance is produced by institutional parameters (Section~2.4).
\end{enumerate}

Yet the central question remains open: why is this movement directional? The
empirical picture introduced at the beginning of the article is asymmetric.
Drift from support toward control is routine, while a return to support is
exceptional. None of the traditions reviewed above fully theorizes this
asymmetry.

This is where the theory of path dependence becomes relevant. Institutional
configurations that generate increasing returns tend to reinforce themselves.
With each cycle of reproduction, staying on the existing path becomes
cheaper, while exit becomes more costly. Reversal then requires an exogenous
intervention whose costs are disproportionate to the costs of the initial
choice (Pierson, 2000; 2004).

A control-oriented configuration of a welfare algorithm is, in ideal-typical
form, a powerful generator of increasing returns. It produces outputs that
can be incorporated into budget justifications and electoral reporting. It
relies on encoded discretion, which cannot be softened through situational
judgment. Once entrenched in statute, it acquires the inertia of a legal
norm.

Within the same framework, a support-oriented configuration lacks all three
of these supports. Its result is counterfactual. Its flexibility depends on
continuous administrative effort. Its legal form is more often permissive
than binding. Under such conditions, routine political pressure
systematically pushes the system toward the control pole and keeps it there.

\begin{figure}[htbp]
\centering
\includegraphics[width=\textwidth]{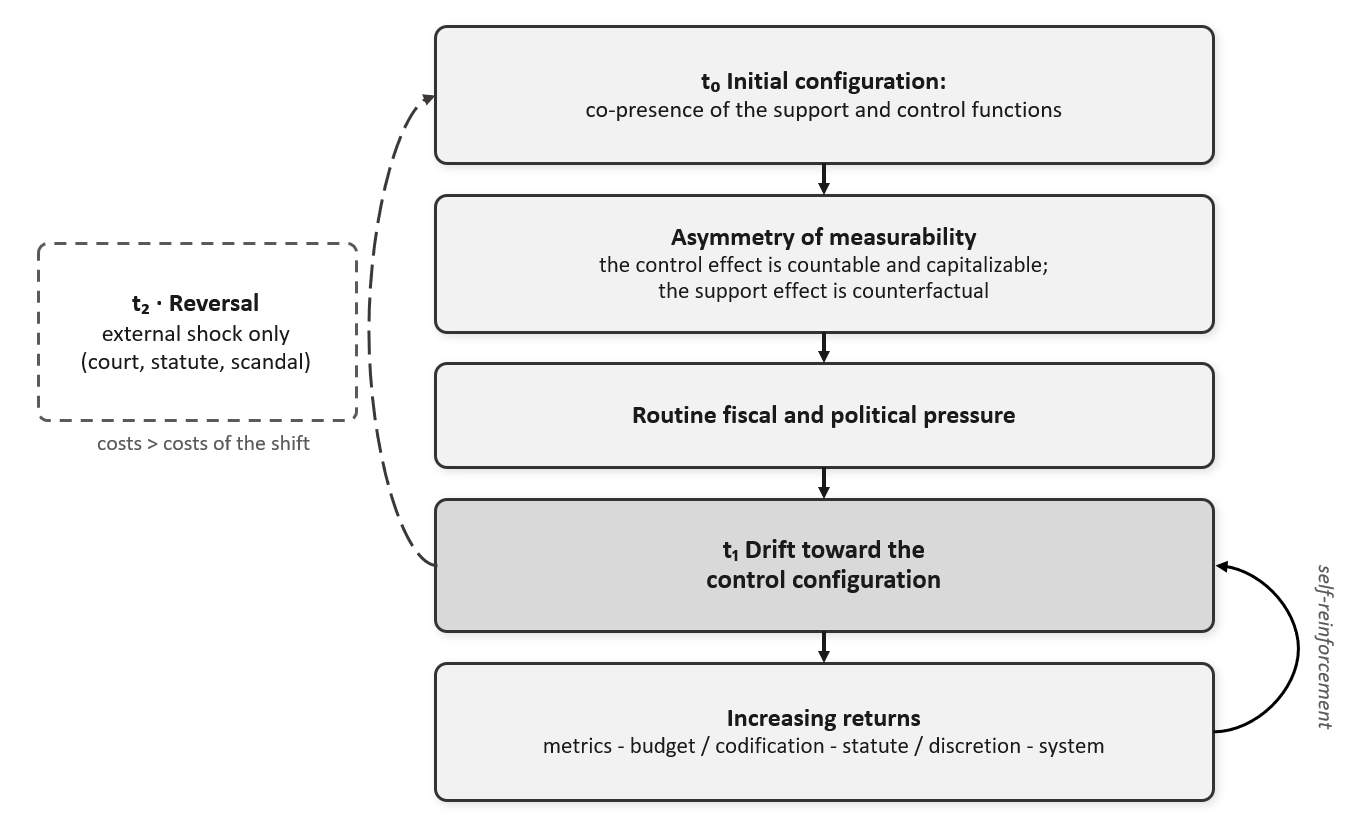}
\caption{\textit{The institutional ratchet: the mechanism of drift from
support to control.} Source: prepared by the author.}
\label{fig:ratchet}
\end{figure}

Movement in the opposite direction is possible, but it follows a different
path: judicial intervention, legislative prohibition, or public scandal. It
is this asymmetry in the available channels that gives drift its
ratchet-like character (Figure~\ref{fig:ratchet}).

The literature reviewed above makes it possible to formulate four
observations that can only be tested empirically.

\medskip
\noindent\textbf{P1. Structural convergence.} In U.S. state and county
welfare administration, the functions of support and control are co-present
within the same system. Their relative balance is shaped by institutional
design, especially by the quality of appeal procedures, the form of legal
codification, and the side on which the system places the costs of error.

\medskip
\noindent\textbf{P2. Directionality of drift.} The balance between these
functions is unstable over time, and its movement is directional: drift from
support toward control is observed systematically more often than movement in
the opposite direction.

\medskip
\noindent\textbf{P3. Asymmetry of measurability.} The directionality of drift
is explained by an asymmetry of measurability. The effects of the control
function are easier to measure and politically capitalize on than the effects
of the support function. As a result, routine political pressure
systematically shifts the system toward control and keeps it in that
configuration.

\medskip
\noindent\textbf{P4. Asymmetry of reversal.} Reverse drift is possible, but
it requires external intervention of disproportionate force, such as judicial
compulsion, legislative prohibition, or public scandal. A return to support
occurs at costs that are an order of magnitude higher than the costs of the
initial shift. This is what gives the asymmetry its ratchet-like character.

\section{Methodology and Data}

The research design follows from the structure of the research questions. The
main question asks: under what institutional conditions does an algorithmic
system in welfare administration acquire a dominant orientation toward
support, and under what conditions does it acquire a dominant orientation
toward control? This is a question about the combination of conditions that
produces an outcome. The derivative question asks whether the transition
between these orientations is symmetrical. This is a question about a
property of the process, one that can only be observed dynamically.

None of the standard ways of organizing comparative qualitative research
answers both questions at once. A most-similar systems design can isolate the
condition that varies across cases, but it captures states rather than
transitions.

The empirical strategy therefore has two stages. The primary method is
within-case process tracing in the sense developed by Beach and Pedersen
(2019). The second component follows the logic of structured, focused
comparison (George \& Bennett, 2005): each case is examined through the same
set of standardized questions, corresponding to the six parameters of the
coding framework. This makes the six case reconstructions comparable. The
evidentiary work on the mechanism takes place within each case, because the
ratchet is a mechanism. Evidence of that mechanism is extracted from
sequences of events, documents, and decisions.

In terms of the type of process tracing, this study is theory-testing, with
an element of scope-condition refinement. Propositions P1--P4 are formulated
deductively at the intersection of four theoretical strands: the historical
sociology of welfare (P1), research on the migration of discretion (P2), the
sociology of quantification (P3), and the theory of path dependence (P4). The
empirical analysis tests this theoretical material. The negative case does
not serve as confirmation. Instead, it helps specify the scope conditions of
the mechanism.

The unit of analysis is an institutional episode in the life of an
algorithmic system: a configuration of six institutional parameters observed
at a particular point in time. If support and control are co-present within
the same system, then it is not systems themselves that should be classified,
but their states.

Each case is reconstructed at a minimum of two, and in some instances three,
time points, depending on whether external intervention occurred:

\begin{itemize}[itemsep=2pt]
\item $t_0$ -- the initial configuration at the moment of implementation;
\item $t_1$ -- the configuration after an episode of institutional pressure;
\item $t_2$ -- the configuration after external intervention.
\end{itemize}

These time points are anchored only to documented institutional events.

\begin{figure}[htbp]
\centering
\includegraphics[width=\textwidth]{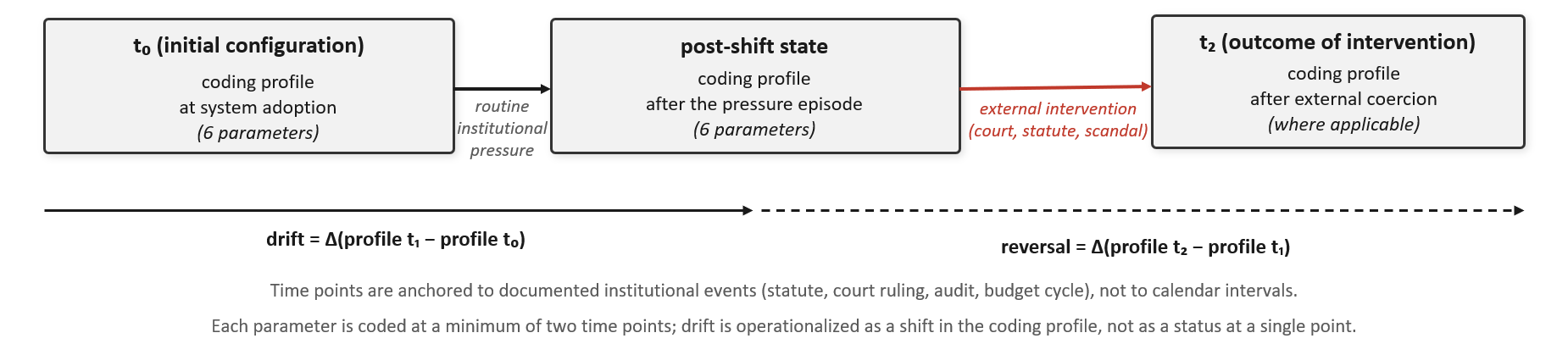}
\caption{\textit{Temporal structure of coding and the operationalization of
drift.} Source: prepared by the author.}
\label{fig:temporal}
\end{figure}

Drift is operationalized as a change in the coding profile between time
points. A system described through six parameters across two or three time
points yields between twelve and eighteen analytical observations. In
Collier's (2011) terms, these are causal-process observations, whose
evidentiary value depends not on their number, but on their position within
the reconstructed sequence. The temporal structure of coding is summarized in
Figure~\ref{fig:temporal}.

\subsection{Case Selection}

The empirical basis of the study consists of six cases of algorithmic systems
in U.S. state and county welfare administration. All six belong to the
subsample of the previous study (Dedyaev, 2026), which ensures continuity in
the empirical field.

The logic of selection follows from the causal role each case plays in the
mechanism under investigation. Each case was selected because it makes a
particular segment of the ratchet observable (Table~\ref{tab:cases}).

\begin{table}[htbp]
\centering
\footnotesize
\begin{tabularx}{\textwidth}{>{\bfseries}p{2.6cm} p{2.6cm} L L}
\toprule
\textbf{Case} & \textbf{Jurisdiction and domain} &
\textbf{Causal role in the mechanism} & \textbf{Key time points} \\
\midrule
NYSDOL AI fraud detection & New York; unemployment insurance &
Routine drift from a crisis instrument to a permanent control
infrastructure without external intervention. &
$t_0$ -- crisis implementation (2020); $t_1$ -- institutionalization in
permanent operations \\
\addlinespace
Michigan MiDAS & Michigan; unemployment insurance &
Drift with forced reversal: the only case in which the full cost of
reversal is observable. &
$t_0$ -- implementation (2013); $t_1$ -- automated fraud determinations
(2013--2015); $t_2$ -- judicial-legislative reversal (2017--2024) \\
\addlinespace
Allegheny Family Screening Tool (AFST) & Allegheny County, Pennsylvania;
child welfare &
Convergence under continuous external audit: a test of the institutional
conditions that prevent the balance between functions from sliding toward
control. &
$t_0$ -- implementation (2016); $t_1$ -- subsequent versions of the tool
and waves of independent evaluation \\
\addlinespace
Illinois Medicaid managed care (HealthChoice Illinois) & Illinois;
Medicaid &
Codification lock-in: the statutory embedding of the algorithm into the
definition of entitlements as a mechanism of irreversibility. &
$t_0$ -- statutory entrenchment of the regime and pilot implementation;
$t_1$ -- statewide implementation \\
\addlinespace
Washington Foster Care & Washington; transition support out of the foster
care system &
Ambivalent drift under fiscal pressure. &
$t_0$ -- pilot implementation; $t_1$ -- configuration under budgetary
constraints \\
\addlinespace
LA County Homelessness Prevention Unit (HPU) & Los Angeles County;
homelessness prevention &
Conditions for drift were present, but drift did not occur; the case helps
define the boundaries of the mechanism. &
$t_0$ -- program launch (2021); $t_1$ -- current configuration
(2022--2025) \\
\bottomrule
\end{tabularx}
\caption{\textit{Cases included in the study and their causal roles.}
Source: prepared by the author.}
\label{tab:cases}
\end{table}

The inclusion of a negative case is a deliberate methodological choice. LA
County satisfies the possibility principle (Mahoney \& Goertz, 2004): drift
toward control could have occurred there, because its typical enabling
conditions were present -- a predictive model built on cross-agency
administrative data, a fiscally strained domain, and the political salience
of homelessness. Its absence is therefore informative: had the mechanism
operated unconditionally, drift should have appeared here as well.

\subsection{Coding Framework}

Each institutional episode is coded across six parameters
(Table~\ref{tab:coding}). The scales are ordinal and contain three to four
criteria. For each parameter, a decision rule specifies what type of
documentary evidence is sufficient for assigning a value.

\begin{table}[htbp]
\centering
\footnotesize
\begin{tabularx}{\textwidth}{>{\bfseries}p{2.4cm} p{4.3cm} L p{2.7cm}}
\toprule
\textbf{Parameter} & \textbf{Scale of values} & \textbf{Decision rule} &
\textbf{Typical sources} \\
\midrule
Declared vs. actual function &
alignment / partial divergence / inversion &
Comparison of the system's programmatic and budgetary justifications with
the observed distribution of decisions and their consequences. &
Program documents, budget requests, audits, appeal statistics. \\
\addlinespace
Quality of appeal procedures &
\textbf{0} -- absent or merely formal; \textbf{1} -- exists, but the burden
and costs fall on the recipient; \textbf{2} -- effective, with sanctions
suspended during review; \textbf{3} -- effective, with the costs of error
internalized by the state &
Evidence of documented practice: the share of revised decisions,
timeframes, allocation of the burden of proof, and the status of benefits
during contestation. &
Court records, administrative and appeal statistics, legal aid reports. \\
\addlinespace
Form of legal codification &
\textbf{0} -- internal administrative practice; \textbf{1} --
administrative regulation; \textbf{2} -- statutory embedding of the
algorithm in the definition of entitlements &
Existence of a legal provision linking the scope or conditions of the
recipient's entitlements to the system. &
Statutes, administrative regulations, legislative materials. \\
\addlinespace
Point and degree of human oversight &
\textbf{0} -- absent or nominal; \textbf{1} -- at the output stage, after
the algorithmic decision; \textbf{2} -- at the input stage, where the human
decision-maker is supported by the system &
Nominality is assessed in practice: documented frequency with which human
reviewers reject algorithmic recommendations. &
Internal rules, independent evaluations, data on the frequency of
recommendation overrides. \\
\addlinespace
Dominant performance metric &
control-oriented: money saved, violations detected / support-oriented:
people reached, harm prevented / mixed &
Which indicators the system presents in budget justifications, public
reporting, and political communication. &
Budget documents, press releases, agency reports. \\
\addlinespace
Distribution of false-positive costs &
borne by the state / mixed / borne by the recipient &
Who bears the material and procedural consequences of system error:
suspension of benefits, recovery of payments, penalties, or, alternatively,
excessive assistance and public expenditure. &
Court records, audits, investigative journalism, appeal statistics. \\
\bottomrule
\end{tabularx}
\caption{\textit{Coding framework: parameters, scales, and decision rules.}
Source: prepared by the author.}
\label{tab:coding}
\end{table}

To ensure that the testing of the propositions does not collapse into mere
illustration, each proposition is paired with an explicit causal test,
following Van Evera (1997) and Collier (2011), and with a pre-specified
falsification condition (Table~\ref{tab:tests}). The tests are understood in
an informal Bayesian sense (Bennett \& Checkel, 2015): passing a test
increases confidence in proportion to how unexpected the observed evidence
would be if the proposition were false.

\begin{table}[htbp]
\centering
\footnotesize
\begin{tabularx}{\textwidth}{>{\bfseries}p{2.7cm} p{2.6cm} L L}
\toprule
\textbf{Proposition} & \textbf{Type of test} & \textbf{Expected evidence} &
\textbf{Falsification condition} \\
\midrule
P1. Structural convergence & Hoop test across all six cases &
In the $t_0$ configuration of each case, both functional components are
present, and the profile of their relative balance corresponds to the
values of the institutional parameters. &
A case with a stable ``pure'' function that does not depend on
institutional design. \\
\addlinespace
P2. Directionality of drift & Pattern test + hoop test &
Among routine transitions from $t_0$ to $t_1$, there are no reversals
toward support. &
A documented routine reversal achieved through managerial adjustment
without external compulsion. \\
\addlinespace
P3. Asymmetry of measurability & Smoking-gun test &
Control metrics are systematically present in budget justifications and
political communication, while supportive effects are either not measured
or not politically capitalized on. &
Systematic public capitalization of prevented harm, comparable in
political weight to the capitalization of savings. \\
\addlinespace
P4. Asymmetry of reversal & Smoking-gun test (MiDAS) + shadow test (other
cases) &
Reversal occurs only through external compulsion of disproportionate
force, and its costs are documented: duration of litigation, volume of
compensation, legislative amendments. &
A case of reversal achieved through a regular managerial procedure, with
costs comparable to those of the initial shift toward control. \\
\bottomrule
\end{tabularx}
\caption{\textit{Correspondence between propositions and causal tests.}
Source: prepared by the author.}
\label{tab:tests}
\end{table}

For each case, the corpus consists of four types of sources: (1) agency and
budget reports produced by the implementing bodies; (2) judicial and
legislative documents, including complaints, court rulings, settlements,
statutes, and hearing materials; (3) investigative journalism and reports by
advocacy and expert organizations; and (4) independent academic evaluations
and audits. The requirement of corpus completeness is defined by the coding
framework: for each case, the matrix ``parameter + time point + source'' must
be filled.

The structure of the codebook follows the format proposed by MacQueen,
McLellan, Kay, and Milstein (1998) for standardized qualitative codebooks.
Each parameter is described through a full definition, anchor definitions for
each value, inclusion and exclusion criteria, a decision rule with a minimum
evidentiary threshold, borderline cases, and typical sources. The procedural
sequence of codebook development follows the iterative protocol proposed by
Fonteyn, Vettese, Lancaster, and Bauer-Wu (2008). Finally, since the coding
is conducted by a single researcher, the codebook performs the dual function
described by Oliveira (2023) for individual coders. In addition to
standardizing the coding procedure, it also serves as a tool for monitoring
the researcher's own consistency.

\subsection{Methodological Limitations}

Several conditions must be specified for this methodology to work. The scope
conditions are inherited from the previous study and refined here.
Accordingly, the generalization offered by this article is analytical rather
than statistical. What is generalized is the mechanism and the conditions
under which it operates.

The study also has three major limitations that should be stated at this
stage. The first is observability bias. The documentary trace on which the
corpus is built is denser where drift became a matter of conflict; quieter
shifts may therefore be underrepresented. The second limitation is the
temporal incompleteness of some cases. In the LA County case, for example, a
key piece of experimental evidence has not yet been published, and the
corresponding codings are therefore marked as preliminary. The final
limitation, partly noted above, is that the coding was conducted by a single
researcher. This is mitigated by the full publication of the codebook.

\section{Results}

The empirical basis of this section consists of six cases reconstructed
across fourteen time points. This produced eighty-four coding cells, each
supported by at least one independent source from a corpus of fifty-two
documents.

\begin{figure}[htbp]
\centering
\includegraphics[width=\textwidth]{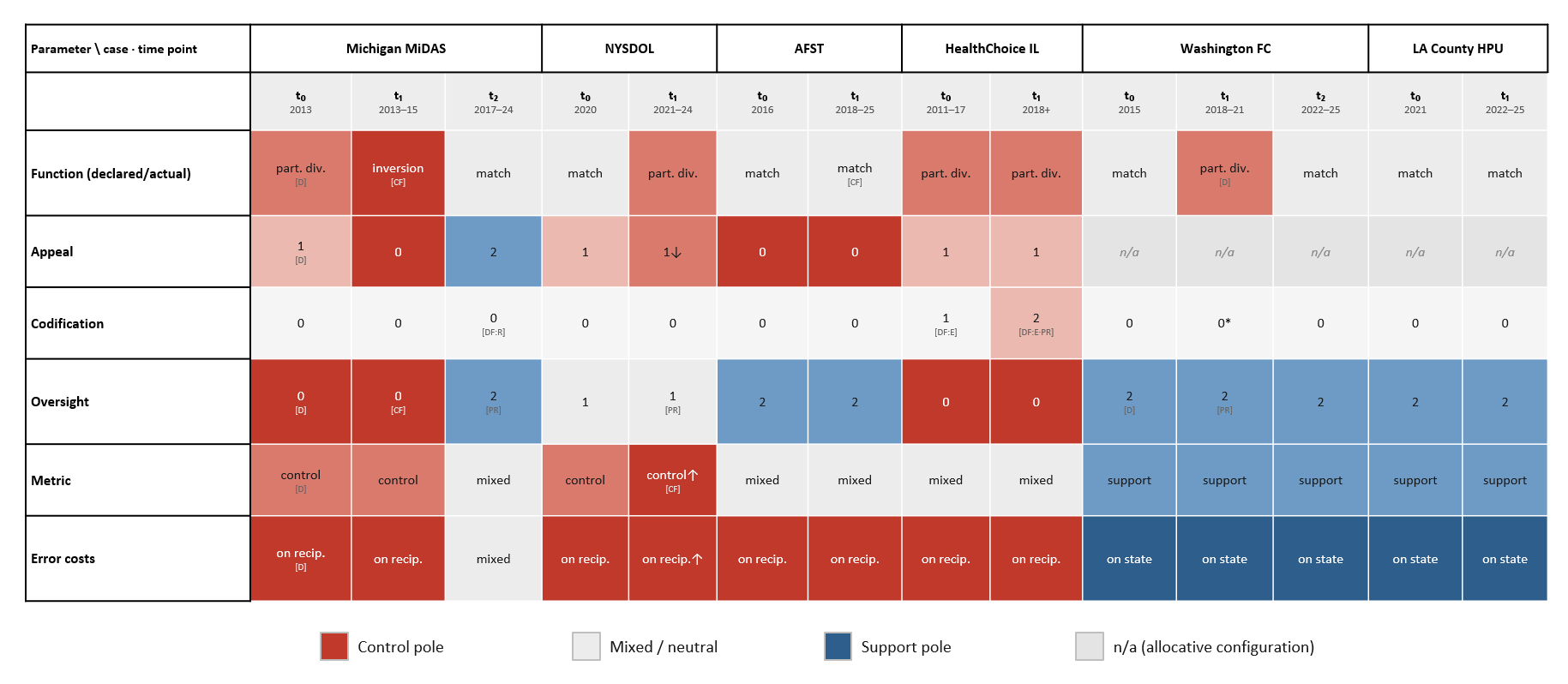}
\caption{\textit{Summary cross-case matrix of coding profiles: six
parameters, six cases, and fourteen time points.} Source: prepared by the
author.}
\label{fig:matrix}
\end{figure}

The summary cross-case matrix (Figure~\ref{fig:matrix}) brings together all
fourteen case configurations. The color of each cell indicates the functional
pole, ranging from support, where the system absorbs the costs of error, to
control, where those costs are shifted onto the recipient. Movement across a
row should be read as the drift of a single parameter over time. Movement
down a column should be read as the configuration of a specific institutional
episode.

\subsection{P1. Structural Convergence}

The results make visible two absences that should be read as findings rather
than gaps. The first is the absence of a third time point in the NYSDOL case.
External intervention did not occur there: the State Comptroller's audit
contained strong findings, but it lacked an enforcement mechanism, while the
relevant bill passed the Assembly but never received a vote in the Senate.

The second is the ``N/A'' value for the appeal parameter in the LA County and
Washington cases. In their support-oriented configurations, these systems do
not produce decisions that are adverse to recipients. A negative system
output does not reduce the scope of entitlements and does not trigger
sanctions. For this reason, there is no object of appeal.

In this context, structural convergence (P1) means that the functions of
support and control are co-present within the same welfare administration
system, and that their relative balance is shaped by institutional design
rather than by the type of model used. The hoop test shows that the strongest
direct evidence of co-presence comes from the Washington case. There, the
decision about which type of error the system is willing to tolerate is
recorded in the design document as a deliberate choice. This provides a
unique form of ex ante predictive evidence for the purposes of this study.

In the other five cases, both components are no less clearly observable at
the moment of implementation. In MiDAS and NYSDOL, the service-oriented
declaration of an emergency lifeline coexists with a blocking architecture: a
system presented as a tool for rapid benefit delivery also contains a
mechanism for automatic suspension. In LA County, the surveillance
infrastructure of profiling is built into the support-oriented system itself.
Co-presence is therefore observable even at the support pole, which
strengthens the hoop test from the opposite side. In HealthChoice Illinois,
allocative and restrictive mechanics (mandatory plan assignment and choice
windows) are embedded in a mandate that is framed rhetorically as care
coordination.

Similar classes of models produce different functional profiles. Predictive
risk scoring underlies AFST, Washington, and LA County, yet these three
systems occupy different poles in the matrix. Conversely, similar profiles
can be built on different technologies: the control-oriented configurations
of MiDAS and NYSDOL rely on technically dissimilar solutions.

\subsection{P2. Directionality of Drift}

The second proposition (P2) states that the balance between functions is
unstable and moves directionally. Drift from support toward control
systematically predominates. This proposition is tested through a pattern
test and a hoop test on the negative case.

The results show that an inventory of routine transitions in the corpus
reveals four distinct types of movement toward control. The first is a
categorical shift involving functional inversion. In the MiDAS case, the
transition from $t_0$ to $t_1$ entails a change in grade on two parameters at
once. The function is inverted: what was declared as support begins to
operate as control. At the same time, the effectiveness of appeal
deteriorates from formally available to practically inaccessible, moving from
value 1 to value 0.

The second type is consolidation without a categorical shift. In the NYSDOL
case, a crisis instrument introduced as an emergency measure becomes, without
any external compulsion, a permanent infrastructure: a four-year
modernization plan with a new detection system, later extended to an adjacent
program. The functional category formally shifts by only one step, but what
becomes entrenched is a control-oriented configuration.

The third type is within-grade degradation. This is where the ordinal scales
provide a methodological advantage. In the NYSDOL case, the quality of appeal
remains coded as \textbf{1} at both $t_0$ and $t_1$. Yet within this
unchanged grade, a substantial deterioration occurs: a prosecutorial
presumption of guilt in the appeal board, fraud determinations without a
requirement to prove intent, and the inaccessibility of live agents under a
self-service priority model. A binary coding of ``appeal exists / appeal does
not exist'' would not capture this shift at all. An ordinal scale, with a
decision rule based on practice rather than formal regulation, captures it as
a real, even if subthreshold, form of drift.

The fourth type is deepening codification. In the HealthChoice Illinois case,
the form of legal codification rises from the first to the second level of
the scale, through the authorizing DF:E flag\footnote{DF:E is a directional
flag on the codification parameter indicating that the legal provision is
enabling -- it authorizes binding a recipient's rights to the algorithm's
output and thereby entrenches the control function.}, while the regime is
simultaneously extended to roughly eighty percent of the state's Medicaid
recipients. At the state level, irreversibility is reproduced separately for
each recipient: silence is constituted as consent to assignment, and the
algorithmically selected plan becomes fixed after the ninety-day choice
window expires.

\subsection{P3. Asymmetry of Measurability}

The third proposition (P3) explains the directionality of drift through an
asymmetry of measurability. The effects of the control function (money saved,
violations detected, and similar outputs) are easier to quantify and to
capitalize on politically. Here, capitalization means the conversion of an
outcome into a publicly presentable achievement, one that can be used in
budget justifications and political communication. The proposition is paired
with a smoking-gun test: the search for a trace of the mechanism that would
substantially increase confidence in the proposition if observed.

The NYSDOL case provides the clearest result. The agency publicly capitalizes
on a figure of more than \$36 billion in ``prevented fraud.'' Yet, when
required to substantiate this figure by the State Comptroller's audit, it was
unable to document it and refused to provide auditors with the data needed
for independent assessment. The metric therefore exists in the channel of
capitalization but not in the channel of verification.

As a factual claim, the reported figure is rejected: it is not confirmed by
an independent source. As evidence of what the agency chooses to place into
public capitalization, however, it is accepted directly. In that capacity, it
constitutes evidence for P3. A cross-corpus check confirms the same pattern.
Control metrics are systematically present in the budgetary and political
channels of MiDAS and NYSDOL: in MiDAS, the growth of the penalty fund is
presented as a public achievement; in NYSDOL, the central figure is prevented
fraud. By contrast, support-oriented indicators are absent from the
capitalized channels in these cases.

Nevertheless, two qualifications should be specified for the subsequent
discussion.

First, the HealthChoice Illinois case demonstrates mixed capitalization
without a clear predominance of punitive metrics. This separates sanctioning
control from allocative control. Based on the corpus, the P3 mechanism
appears to be specific to configurations that impose sanctions; it does not
extend to control as such.

Second, in LA County, in the second phase of AFST, and in Washington, the
support function gains a measurement and capitalization infrastructure where
such infrastructure is deliberately built. In LA County, this role is played
by an affiliated research laboratory. In the second phase of AFST, the corpus
documents a shift in capitalization toward support-oriented indicators. In
Washington, the statutory goal is linked to a reporting metric for reducing
homelessness. Where this infrastructure exists, prevented harm can be
measured and publicly presented. The asymmetry proposed in P3 should
therefore be located in routine budgetary and political channels, rather than
in the intrinsic nature of support-oriented effects.

\subsection{P4. Asymmetry of Reversal}

The fourth proposition (P4) states that reverse drift is possible, but that
it requires external intervention of disproportionate force, and that its
costs are an order of magnitude higher than the costs of the initial shift.
The proposition is tested through a smoking-gun test in the MiDAS case and a
shadow test across the remaining cases.

\begin{figure}[htbp]
\centering
\includegraphics[width=\textwidth]{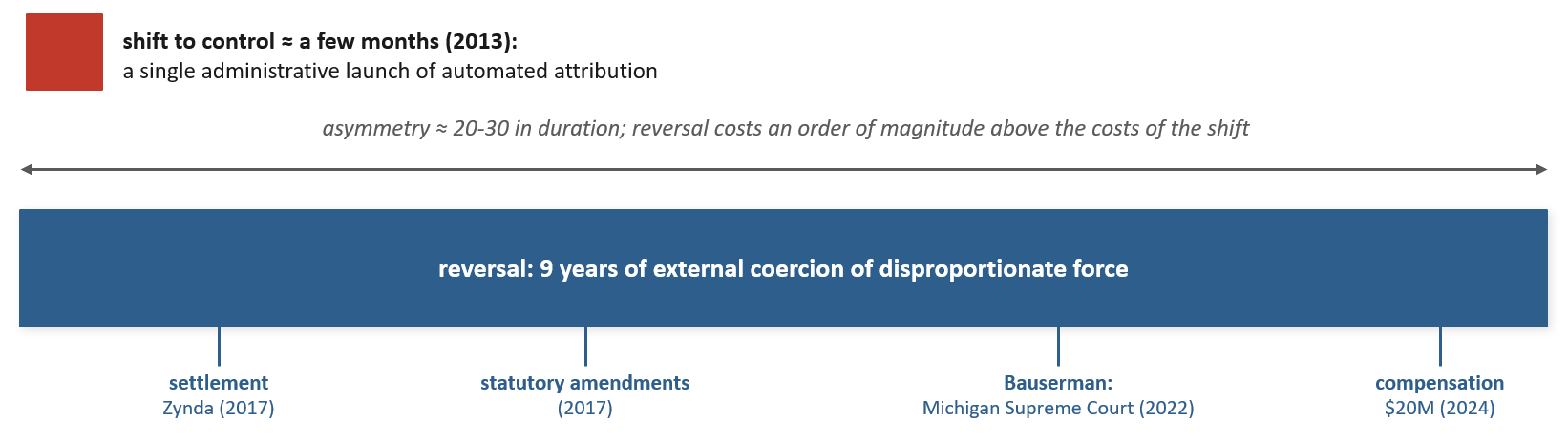}
\caption{\textit{The asymmetry of reversal costs in the MiDAS case: the
administrative ease of the initial shift versus a nine-year sequence of
external interventions.} Source: prepared by the author.}
\label{fig:midas}
\end{figure}

The results show that MiDAS occupies the position of a test bed in the
overall design, because the full cost of reversal is observable and
procedurally documented in this case. The automated attribution of fraud to
roughly forty thousand recipients was initiated by a single administrative
decision over several months in 2013. It was stopped only through a sequence
of external interventions that extended over nine years: the settlement in
\textit{Zynda} (2017), statutory amendments adopted the same year, the
Michigan Supreme Court's decision in \textit{Bauserman} (2022), which
recognized the possibility of recovering constitutional damages, and the
court-approved \$20 million compensation settlement in 2024. The asymmetry in
duration and cost is visible directly in this chronology
(Figure~\ref{fig:midas}).

A separate result should also be registered: even forced reversal does not
restore the original configuration. At $t_2$, the function is coded as
``alignment,'' but this alignment is achieved through the legalization of the
control function. The codification parameter receives the first DF:R flag in
the corpus -- a restrictive directional flag, the mirror of DF:E, marking
that the new legal provision constrains rather than authorizes the binding of
a recipient's rights to the algorithm's output -- while the costs of error
remain mixed.

The Washington case provides a useful contrast. It includes two external
interventions -- SSB 6560 (2018) and HB 1905 (2022). Both were legislative,
both moved in a support-oriented direction, and both were relatively low-cost
and repeatable: two acts adopted through the ordinary state legislative
process within four years. Moreover, the second act provided additional
funding for the support-oriented configuration rather than converting it into
cheaper control. The contrast is therefore visible not only within MiDAS, but
also across cases: reversing control is costly, while reinforcing support can
be comparatively cheap.

The remaining cases clarify the threshold of external pressure. NYSDOL shows
that critical external signals do not change the configuration if they lack
coercive force. AFST shows the same pattern. Nine years of external pressure
(two waves of commissioned independent evaluation, an external academic
audit, a national journalistic investigation, and federal scrutiny of the
system's civil-rights profile) did not move a single parameter. The resulting
observation is that the level of external force required to shift an
established algorithmic configuration is high in both directions. The corpus
contains no case of reversal achieved through an ordinary managerial
procedure at comparable cost.

\subsection{Conditions Blocking Drift}

The results section also warrants separate attention to the blocking
conditions that prevented drift from occurring.

The LA County case is especially instructive. The conditions that could have
enabled drift are documented by a high-quality independent source: according
to the California State Auditor, the state spent \$24 billion on homelessness
between 2018 and 2023 while the number of people experiencing homelessness
continued to rise. This made the domain politically and fiscally sensitive.
Yet over four years, none of the six parameters shifted toward control.

The coding makes it possible to identify four specific blockers:

\begin{itemize}[itemsep=2pt]
\item The costs of error are structurally internalized by the state. The
      system simply has no sanctioning output through which those costs
      could be shifted onto the recipient.
\item Human judgment is positioned at the input stage. The algorithmic
      output does not have the status of a decision.
\item There is no legal attachment. The model has nothing through which it
      can become locked into the definition of entitlements.
\item A regime of continuous external evaluation is in place. The system is
      voluntarily subjected to experimental assessment.
\end{itemize}

The Washington case adds a fifth blocker: the statutory codification of a
support-oriented goal, to which the legislature returned the system twice.
This reveals a form of symmetry that was not obvious from the theoretical
framework. In the corpus, codification mainly appears as a mechanism of
control lock-in, as in Illinois. At the support pole, however, it works as a
mechanism for preserving support. The same parameter can therefore entrench
either pole.

The AFST case protects the result from being read as ``blocking equals
welfare-enhancing.'' Nine years of operation under the densest external
pressure in the sample did not shift the configuration. Yet what remained
stable was not only the score. Human oversight at the input stage also
remained stable -- the only such case in the corpus confirmed by published
data on the frequency of recommendation overrides. So did the absence of
appeal and the distribution of error costs. The configuration prevents the
balance between functions from sliding toward control, but it also preserves
the price of that balance.

\begin{figure}[htbp]
\centering
\includegraphics[width=\textwidth]{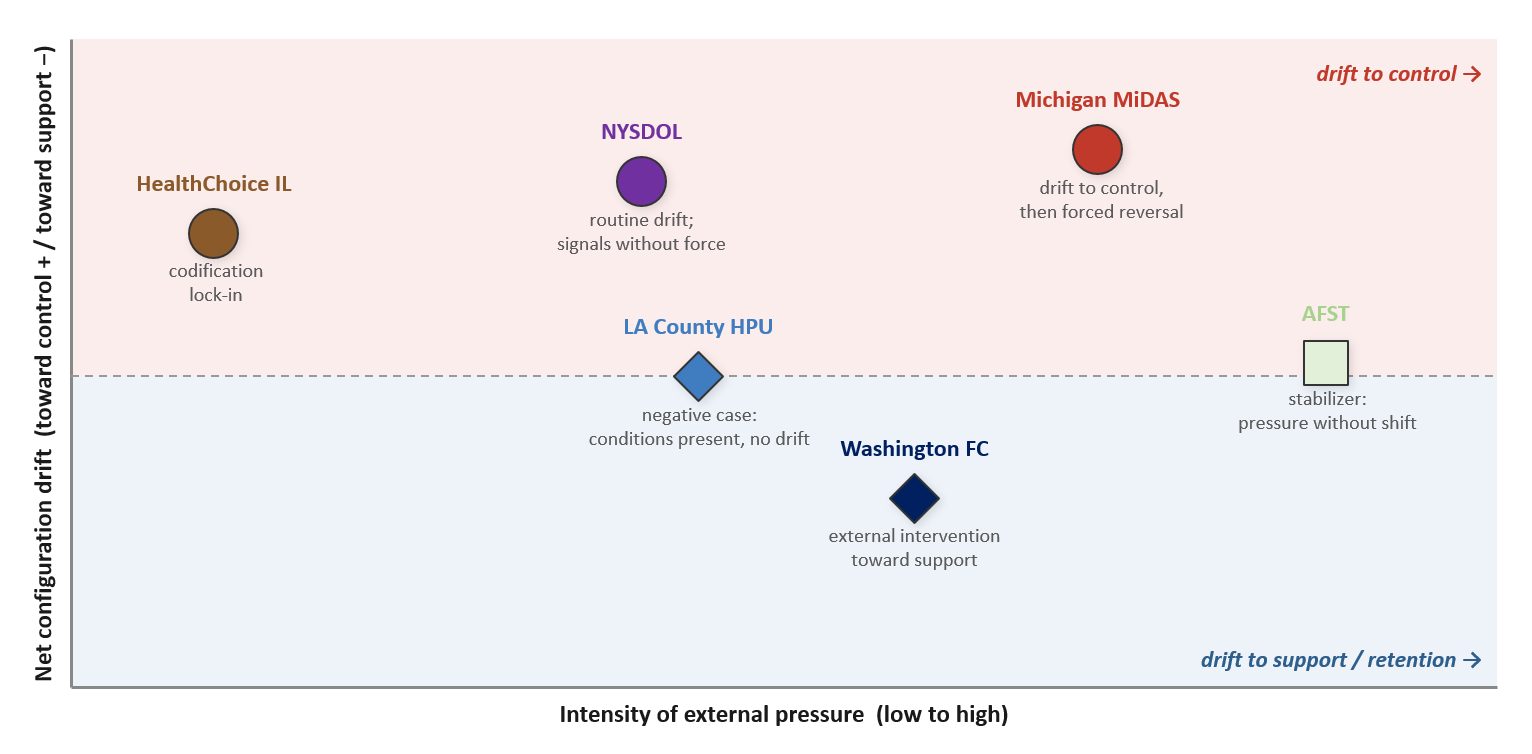}
\caption{\textit{External pressure and net drift: cases positioned by the
intensity of external pressure and the direction of the resulting shift.}
Source: prepared by the author.}
\label{fig:pressure}
\end{figure}

Across the corpus, the intensity of external pressure and the net drift of
the configuration are not correlated (Figure~\ref{fig:pressure}). AFST
remains immobile under maximum pressure. Illinois drifts with almost no
external pressure. Direction is set by the configuration, not by the strength
of pressure.

\subsection{Alternative Explanations}

A party-ideological explanation would predict that shifts cluster in
jurisdictions under a particular form of partisan control. The observed
pattern is different. Drift toward control is found in New York and Michigan
under different and changing configurations of partisan control. At the same
time, drift is absent in Democratic California and Washington, while a stable
control-oriented configuration, with an unchanged distribution of costs,
persists in the equally Democratic Allegheny County. The predicted clustering
is therefore absent. Six cases are not sufficient to reject the partisan
hypothesis in a statistical sense. The test is conducted in the logic of a
pattern test: the regularity predicted by the literature is not observed in
the corpus.

A vendor-based explanation would attribute the direction of drift to the
business logic of the technology provider. Yet the corpus includes different
types of developers: an external contractor in MiDAS, in-house agency
development in NYSDOL, and academic or research teams in AFST, the California
Policy Lab, and RDA in partnership with United Way. The ways do not cluster
by provider type.

A crisis-based explanation would reduce drift to a response to shock-induced
overload. This explanation is built into the design through the NYSDOL-MiDAS
pair: both unemployment insurance systems emerged from the pandemic crisis,
but they diverged across institutional parameters. LA County shows the
absence of drift outside a context of crisis overload, while Washington shows
the stability of support under fiscal pressure. Crisis should therefore be
treated as a condition that accelerates the mechanism, not as one that
produces it.

\section{Discussion}

\begin{figure}[htbp]
\centering
\includegraphics[width=\textwidth]{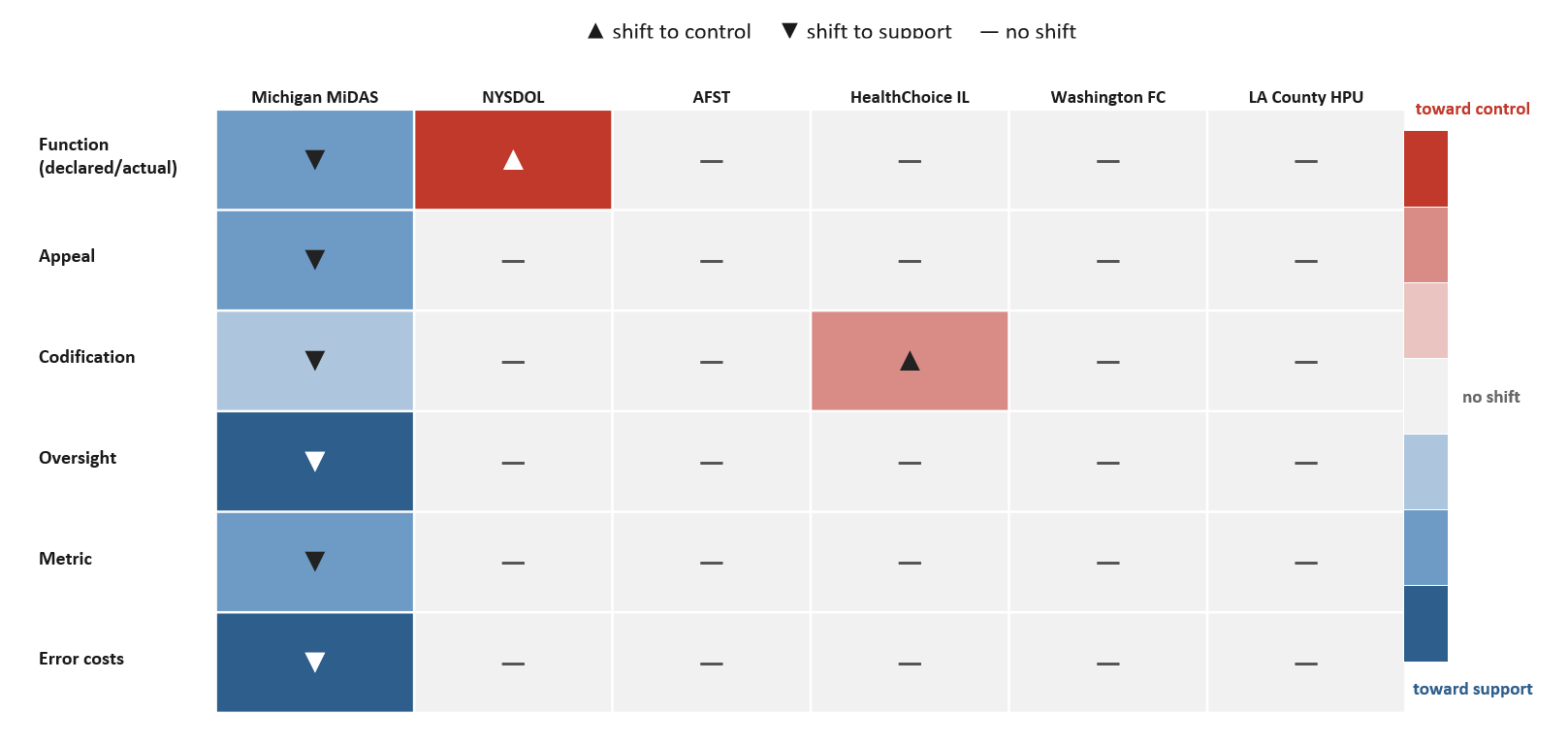}
\caption{\textit{Direction and presence of drift across parameters and cases:
what moves the ratchet.} Source: prepared by the author.}
\label{fig:channels}
\end{figure}

The discussion returns to the gap that motivated the article. In the previous
typology, the state level was captured by the residual category of the
``hybrid'' (Dedyaev, 2026). In light of the present corpus, that category
becomes a family of trajectories. Drift is concentrated in three parameters:
function, appeal, and codification. By contrast, human oversight and the
distribution of error costs remain stable in established configurations. They
operate as upstream blockers of drift (Figure~\ref{fig:channels}).

The observed pattern (the directionality of routine drift, the asymmetry of
reversal costs, and the identifiable blockers) is consistent with the
mechanism of the institutional ratchet.

Reconstructing the mechanism adds two elements that were not fully specified
in either the theoretical framework or the methodology.

The first is a specification of the roles played by the parameters. Drift
proceeds through three channels: function, appeal, and codification. By
contrast, the point of oversight and the distribution of error costs remain
stable in established configurations. This leads to an important narrowing of
the argument. The ratchet is not a property of welfare AI as a class of
technologies. It is a property of a specific institutional configuration: one
that combines a sanctioning output, oversight at the output stage or no
meaningful oversight at all, and escalating codification. Outside this
configuration, the mechanism is not activated.

The second element is that forced reversal does not return the system to its
original point. Instead, it legalizes the control function in a new legal
form while leaving the costs of error mixed. The ratchet has memory: even
under compulsion, the machine is re-entrenched. Together with the threshold
observation, that an established configuration requires a high level of
external force to move in either direction, this yields the final formulation
of the mechanism: routine politics supplies a one-directional pressure; the
configuration determines whether that pressure is converted into movement,
and reversal can be purchased only outside the routine circuit.

The findings also clarify how the results relate to the theoretical
framework. The historical sociology of welfare is confirmed in its core
premise, but revised in its account of dynamics. The co-presence of
assistance and discipline is indeed constitutive of the welfare state (Piven
\& Cloward, 1971). Yet the cyclical symmetry in which benefits expand during
periods of instability and tighten during periods of order does not survive
algorithmic codification. The expansionary phase remains politically
available only until the configuration is fixed in code and statute. After
that, the regulation of poverty becomes one-directional at the level of
routine administration.

The framework of function creep (Koops, 2021) is also supplemented rather
than rejected. In the NYSDOL case, function creep (the extension of the tool
to the Excluded Workers Fund) occurs alongside within-grade degradation of
appeal procedures under an unchanged mandate. These two processes are
simultaneous, but they are coded through different parameters. This shows
that the conceptual space occupied by support--control convergence was not
already covered by function creep.

The corpus most strongly revises the sociology of quantification. The claim
about the reactivity of measurement is confirmed (Espeland \& Sauder, 2007;
Bevan \& Hood, 2006). However, the asymmetry of measurability turns out to be
an institutionally produced relationship. Where a measurement infrastructure
for prevented harm is deliberately built, support-oriented effects can be
measured and publicly presented. Metric reactivity is therefore configurable.

The findings also refine the literature on human oversight. Green's (2022)
claim about the largely nominal character of oversight holds for oversight at
the output stage. But the position of oversight matters. In AFST, input-stage
oversight, verified by published data on the frequency of recommendation
overrides, works as a blocker of drift rather than as a legitimizer of the
system's output.

Finally, the results partly revise the theory of path dependence. Increasing
returns do not belong naturally to the control pole. Codification can
entrench both poles: control in Illinois and support in Washington. What is
asymmetric, then, is not the mechanism of lock-in itself, but what routine
politics feeds into that mechanism (Pierson, 2000).

\section{Conclusions}

The central conclusion of this article is that it closes the gap identified
at the outset and answers the research question: \textit{under what
institutional conditions does an algorithmic system in welfare administration
acquire a dominant orientation toward support or toward control, and is the
transition between these orientations symmetrical?}

The results show that the direction of movement is shaped not by the
technical model itself, but by a political configuration of six institutional
parameters. Among these, the distribution of false-positive costs is
decisive. The transition between support and control is asymmetric. Routine
politics systematically pushes systems toward control through three channels:
function, appeal, and codification. Movement in the opposite direction
requires external compulsion of disproportionate force. Even then, the system
does not return to its original state, but enters a third configuration.

Drift, however, can be blocked. The ratchet can be stopped before it starts
when several conditions are in place: the structural internalization of error
costs by the state, human judgment at the input stage, the absence of legal
attachment, a regime of continuous external evaluation, and a statutory
anchor for a support-oriented goal.

The contrast is easiest to see in two systems that were built in broadly
similar ways but ended in opposite outcomes. In Michigan, MiDAS was launched
to identify people fraudulently receiving unemployment benefits. The system
was configured so that, at the slightest suspicion, it could designate a
recipient as a fraudster automatically, without human involvement, and demand
repayment with a substantial penalty added. Within a few years, it falsely
accused roughly 40,000 people. Turning this machine on required a single
administrative decision and a few months. Turning it off required nine years
of litigation, a state Supreme Court decision, and a \$20 million
compensation settlement for those harmed.

In California, Los Angeles County adopted a system of the same broad type:
risk prediction based on data from multiple agencies. But the central
institutional choice was made differently from the start. If the system made
an error, the cost of that error would fall on the state: someone might
receive extra assistance they may not ultimately have needed, rather than
being wrongly denied support or punished. As a result, no slide toward
sanctioning occurred. The difference lay in one decision made before
implementation: on which side to place the cost of error.

The article makes contributions at three levels:

\begin{itemize}[itemsep=3pt]
\item \textbf{Conceptual.} The residual category of the ``hybrid,'' which
      the previous typology used to describe the state level (Dedyaev,
      2026), becomes a family of trajectories. Static ``hybridity'' is
      revealed to be a snapshot of movement at the moment of observation.
\item \textbf{Theoretical.} The institutional ratchet is translated into a
      tested mechanism. The four propositions, formulated at the
      intersection of four literatures, passed the pre-specified causal
      tests.
\item \textbf{Methodological.} The coding framework, with ordinal scales,
      decision rules, and a full audit trail, is reproducible beyond the
      sample. It turns the debate over ``care or control'' into measurable
      system behavior.
\end{itemize}

For decades, the field has asked what a welfare algorithm is: an instrument
of care or an instrument of control? Both traditions gave correct answers to
a poorly framed question. The corpus shows that welfare AI is neither one nor
the other, because it moves. It is a political process in a technical shell.
The ratchet turns not because algorithms naturally gravitate toward
surveillance, but because routine budget politics can count money saved and
cannot count crises that never occurred, unless the state builds an
institutional mirror for prevented harm. Where such a mirror is built, the
ratchet stops. This leads to a practical principle that extends beyond
welfare: reversibility is a crucial design property, one that is cheap to
build in at the input stage and an order of magnitude more expensive to
recover at the output stage.

Several trajectories for further research follow from this argument. The
first is defined by the main limitation of the research design. A corpus
built from documentary traces underrepresents quiet shifts. Studying such
shifts would require a different empirical strategy, including requests for
administrative data and systematic monitoring of administrative regulation
across jurisdictions. The second direction is to test the mechanism beyond
the state level. Supranational regulatory regimes that codify algorithms
through risk classes offer a natural test of whether the ratchet depends on
the institutional pluralism of a decentralized system or whether it can also
be reproduced in a unitary setting. Finally, the findings on blockers open a
third direction. If blockers can be identified substantively, the next
logical step is to study the conditions under which reversibility can be
built into an algorithmic system as both a statutory and architectural
requirement.

These findings show that the ``hybrid'' state level was not a conceptual
residue. It was a static image produced by an instrument not designed for
moving objects. This study has built an instrument suited to such objects.
The field now has a tested mechanism for explaining the direction in which
welfare AI systems move at the state level in the United States.


\end{document}